\documentclass[floatfix,11pt,onecolumn,showpacs,amsmath,amssymb]{revtex4}

\usepackage{epsf}
\usepackage{graphicx}  
\usepackage{dcolumn}   
\usepackage{bm}        
\usepackage{color}

\newcommand{\be}{\begin{equation}}
\newcommand{\en}{\end{equation}}
 \newcommand{\bea}{\begin{eqnarray}}
 \newcommand{\ena}{\end{eqnarray}}
  \newcommand{\sch}{Schwarzschild}

\begin{document}

\title{Towards a sound massive cosmology}
\author{Hongsheng Zhang $^{1,~4} $\footnote{Electronic address: sps\_zhanghs@ujn.edu.cn}, Ya-Peng Hu $^{2,~4}$ \footnote{Electronic address: huyp@nuaa.edu.cn}, Yi Zhang$^{3~}$ \footnote{Electronic address: zhangyia@cqupt.edu.cn}}
\affiliation{$^1$ School of Physics and Technology, University of Jinan, 336 West Road of Nan Xinzhuang, Jinan, Shandong 250022, China\\
$^2$ College of Science, Nanjing University of Aeronautics and Astronautics, Nanjing 210016, China\\
$^3$ College of  Science, Chongqing University of Posts and Telecommunications,  Chongqing 400065, China \\
$^4$ State Key Laboratory of Theoretical Physics, Institute of Theoretical Physics, Chinese Academy of Sciences, Beijing, 100190, China}


\begin{abstract}
 It is known that  de Rham-Gabadadze-Tolley (dRGT) massive gravity does not permit a homogeneous and isotropic universe with flat or spherical spatial metrics. We demonstrate that a singular reference metric solves this problem in an economic and straightforward way.
 In the dRGT massive gravity with a singular reference metric, there are sound homogeneous and isotropic cosmological solutions. We investigate cosmologies with the static and dynamical singular reference metrics, respectively. The term like dark energy appears naturally and the universe accelerates itself in some late time evolution.  The term simulating dark matter also naturally emerges. We make a preliminary constraint on the parameters in the dRGT massive gravity in frame of the present cosmological model by using the data of supernovae, cosmic microwave back ground radiations, and baryonic acoustic oscillations.

\end{abstract}


\preprint{arXiv: }
 \maketitle

\section{Introduction}

   In modern field theory, a free field is described by its propagator in a back ground spacetime. It is easy to introduce a mass for a scalar field $\phi$, for which a mass leads to a term $\phi^2$ in the action. Also, it is not difficult to introduce a mass term for a vector field $A_{\mu}$, which leads to a term $A_{\mu}A^{\mu}$ in the action. Unexpectedly, to endow a mass for a spin-2 tensor field is a highly non-trivial topic. In the other view, it is not surprised since gravity is special in several aspects. The spin-2 tensor field also weaves the spacetime background for its own to propagate.  Even we do not consider such a complexity, to construct a massive graviton propagating in the Minkowski spacetime is not a trivial problem. In the linear limit general relativity (GR) in vacuum space becomes a theory of a free spin-2 tensor field $\eta_{\mu\nu}$, where only kinetic terms of the graviton appear. Mimicking the case of scalar and vector fields, one may introduce a mass for the spin-2 tensor field $h_{\mu\nu}=g_{\mu\nu}-\eta_{\mu\nu}$ through a term,
     \be
     m^2 g_{\mu\nu}g^{\mu\nu},
     \label{mgg}
     \en
     in the Lagrangian.   It is clear such a term does not imply a massive graviton (However, see \cite{liuliao}). In Minkowski field theory, it is a shift of vacuum energy. In GR, it is a cosmological constant. In 1939, Fierz and Pauli found the proper massive linear GR \cite{FP1},
     \be
     S=\int d^4x \left(-\frac{1}{2}\partial_{\alpha}h_{\mu\nu}\partial^{\alpha}h^{\mu\nu} +\partial_{\mu}h_{\nu\alpha}\partial^{\nu}h^{\mu\alpha}-\partial^{\mu}h_{\mu\nu}\partial^{\nu}h +\frac{1}{2}\partial_\alpha h  \partial^\alpha h          \right)+ \left(\frac{1}{2} m^2 (h^2-h^{\mu\nu}h_{\mu\nu})\right),
     \en
     where $h$ denotes the trace of the spacetime fluctuation $h_{\mu\nu}$. One easily recognizes that the terms in the first bracket in the above action is the linearized Einstein-Hilbert term. The improvement of  Fierz and Pauli is to introduce the mass term in the second bracket. The critical property of this term is that the relative coefficient of the terms $h$ and  $h^{\mu\nu}h_{\mu\nu}$ is $-1$, which eliminates the ghost freedom. The linearised massive gravity suffers from a non-continuity problem, which can be recovered through the Vainshtein mechanism \cite{vain}.  However, when one generalizes this linearized massive gravity to the non-linear regime, the ghost freedom will reemerge, which is called Boulware-Deser ghost \cite{BDghost}. A satisfactory non-linear ghost-free massive gravity is only recently proposed in \cite{drgt}, in which the property of ghost-free is obtained by successive expansion to the 4th order. Then Hassan et al makes a complete demonstration  that dRGT massive gravity is ghost-free via a Hamiltonian approach \cite{hassan}. It is instructive to put the massive gravity into the large frame of modified gravities, in which several theories permit massive modes. In \cite{massivewave}, the massive modes of higher-order gravity is studied. And the possible observation effects of the massive modes at the upcoming gravitational detectors, especially LISA, are explored. In a review of modified gravity \cite{review}, the general condition on massive terms can emerge in the effective action of gravities is investigated, and related cosmologies is reported. This presents a general frame to find massive modes in modified gravities. Stability problem of Lorentz breaking massive gravity in spherically symmetric spaces is explored in \cite{lbmassive}. The  Vainshtein mechanism in the scale of clusters of galaxies is studied in \cite{vaigala}.

     The recently renewed interest of massive gravity is motivated, more or less, by the cosmic acceleration,  which is a significant discovery over last century. If the universe only contains matters like terrestrial matters described by the standard model, the cosmic expansion must be decelerating rather than accelerating. Many models for this acceleration has been proposed. However, although fundamental for
    our understanding of the universe, its nature, especially in the
        theoretical aspect, remains a completely open question nowadays. The heuristic argument of the possible mechanism of the cosmic acceleration in frame of massive gravity is as follows. Generally, a massless intermediate boson leads to a Newton-like potential $1/r$. A massive intermediate boson leads to a Yukawa-like potential $e^{-\alpha r}/r$, which implies a weakened force at large distance. Thus, for a massive graviton, the gravitational force is weakened, and the universe becomes to be accelerated at large distance. This argument also sets the mass scale of the graviton $m\sim \alpha\sim H_0$, where $H_0$ is the Hubble constant. It seems that a weakened gravity still cannot explain the accelerated universe. One gets inspirations from the ``raw" massive gravity (\ref{mgg}), which implies that the massive cosmology may share some property of the de Sitter universe. Unexpectedly, a technical detailed investigation of the refined massive gravity (dRGT) shows that a homogeneous and isotropic (FRW) universe with flat and spherical 3 spatial spaces is prohibited in dRGT massive gravity \cite{drgtcos}, said nothing of accelerated universe. To solve this problem is the main goal of the present article.

        This article is organized as follows. In the next section, we discuss the difficulties of dRGT massive cosmology, and demonstrate that a singular reference metric yields  sound FRW cosmological solutions. In section III, we study the massive cosmology with singular reference metric in detail. In section IV, we fit this model by using observation data of SNe Ia, CMB,  and BAO. Section V concludes this article.

     \section{difficulty and way out of dRGT massive cosmology}

      In field theory, mass term is a special potential term, which is a scalar in the Lagrangian. To construct a potential for the metric $g_{\mu\nu}$ is not difficult. In fact, $g_{\mu\nu}g^{\mu\nu}$ up to a factor is the unique one which can be constructed directly from $g_{\mu\nu}$, without derivatives of $g_{\mu\nu}$. As we have mentioned, such a term is always a constant, and thus can not be treated as the mass term for gravitons. We name this constant-potential problem. Thus, in order to construct a proper scalar invariant as the mass term it is necessary to introduce some auxiliary fields, for example two auxiliary vectors or an auxiliary tensor. One can use the metric $g_{\mu\nu}$ together with the auxiliary fields to construct a proper potential for the gravity field $g_{\mu\nu}$, which escapes the constant-potential problem. Such an  auxiliary tensor is successfully found in \cite{drgt}, which is dubbed reference metric.
             In the original dRGT model, the reference metric is set to be the Minkowski metric in the inertial frame $f_{\mu\nu}$=diag($-1,1,1,1$). An obvious problem is that it becomes non-covariant.  The solution is to introduce four Stuckelberg scalar fields $\phi^{a}$, and sets,
       \be
       f_{\mu\nu}=\bar{f}_{ab}\frac{\partial \phi^a}{\partial x^{\mu}}\frac{\partial \phi^b}{\partial x^{\nu}},
        \label{ref}
        \en
       where $\bar{f}_{ab}$ is the inner metric in the field space. Thus $f_{\mu\nu}$ becomes covariant with the help of the derivative operators. Armed with the reference metric, the general dRGT potential can be written as,
       \be
       V(g,f)=m^2 \sum^4_i c_i {\cal U}_i (g,f),
        \label{poten1}
        \en
          where,
          \bea
\label{poten}
&& {\cal U}_1= [{\cal K}], \nonumber \\
&& {\cal U}_2=  [{\cal K}]^2 -[{\cal K}^2], \nonumber \\
&& {\cal U}_3= [{\cal K}]^3 - 3[{\cal K}][{\cal K}^2]+ 2[{\cal K}^3], \nonumber \\
&& {\cal U}_4= [{\cal K}]^4- 6[{\cal K}^2][{\cal K}]^2 + 8[{\cal K}^3][{\cal K}]+3[{\cal K}^2]^2 -6[{\cal K}^4].
\ena
       The new tensor $\cal K$, which plays the role of $g_{\mu\alpha}g^{\alpha\nu}$ in the raw massive gravity, is defined as,
       \be
       {\cal K}^{\mu}_{\nu}= \left(\sqrt {g_M^{-1}f_M}\right) ^{\mu}_\nu.
       \label{root}
       \en
         Here $g_M^{-1}$ denotes $g^{\alpha\beta}$ in matrix form, and $f_M$ denotes $f_{\alpha\beta}$ in matrix form. $[{\cal K}]$ labels the trace of ${\cal K}$  measured by
         the spacetime metric $g_{\mu\nu}$. $c_i$ are four constant. A special note is that the root operation of a matrix is a complex problem.  For example the most simple matrix diag(1,1) has at least the following different square roots: the three Pauli matrices, diag(1,1), diag(1,-1), diag(-1,1), and diag(-1,-1). So the operation of square root for a matrix is alive with ambiguity. We take the real matrix with Lorentzian signature as the proper square root in (\ref{root}). With the potential (\ref{poten1}) involving the reference metric, the full action of a massive gravity system reads,
         \be
\label{action}
S =\int d^{4}x \sqrt{-g} \left[ \frac{1}{2\kappa^2}\left(R +V(g,f)\right)+{\cal L}_m\right],
\en
 where $R$ is the Ricci scalar, ${\cal L}_m$ denotes the Lagrangian of the matter fields.

        Interestingly, it is found that the above construction of dRGT does not permit a homogeneous and isotropic (FRW) universe with flat or spherical spatial metrics \cite{drgtcos}. To highlight where the crux lies, we show some technical details of this problem. We take the spatially flat FRW universe as an example. The spatially flat FRW metric reads,
       \be
       ds^2=-dt^2+a^2d\vec{x}^2,
       \label{FRW}
       \en
       where $a$ denotes the scale factor. We set the Stuckelberg fields as,
       \be
       \phi^0=u(t),~\phi^i=x^i.
       \en
     The reference metric is calculated by (\ref{ref}),
     \be
     f_{00}=-\dot{u}^2,~f_{ii}=1,
     \en
      in which we have set $\bar{f}_{ab}=\eta_{ab}$. Then the potential (\ref{poten1}) reads,
      \be
      V=m^2\left(c_1 (\dot{u}+\frac{3}{a})+ c_2 (\frac{6\dot{u}}{a}+\frac{6}{a^2}) +c_3 (\frac{6}{a^3}+\frac{18\dot{u}}{a^2})+ c_4\frac{24\dot{u}}{a^3}\right).
      \label{poten2}
      \en
      The invariant potential in the action,
      \be
      \sqrt{-g}~V= m^2\left(\dot{u}(c_1a^3+6c_2a^2+18c_3a+24c_4)+3c_1a^2+6c_2a+6c_3\right).
            \en
      The terms other than $V$ in the action does not contain $u$. Then a variation with respect to $u$ presents,
      \be
      \frac{d}{dt}(c_1a^3+6c_2a^2+18c_3a+24c_4)=0,
      \label{constr}
      \en
       which implies that $c_1a^3+6c_2a^2+18c_3a+24c_4$ is a constant, and thus $a$ is a constant in the history of the universe. Only a static universe is permitted.
       The other point of view of this problem is to work in the unitary gauge, i.e., $\phi^a=\delta^a_\mu x^\mu$. Because of this gauge fixing, one should introduce
       the lapse function in the ADM decomposition in the FRW metric (\ref{FRW}),
       \be
       ds^2=-N^2(t)dt^2+a^2d\vec{x}^2.
       \label{FRWp}
       \en
      To obtain ${\cal U}_1,~{\cal U}_2,~{\cal U}_3,~{\cal U}_4$, we need only replace $\dot{u}$ by $N$. The resulted action reads,
      \be
      S=\int d^4x \left(A(a,~\dot{a})+Nm^2(c_1a^3+6c_2a^2+18c_3a+24c_4)\right),
      \label{multiplier}
      \en
  where $A(a,\dot{a})$ is some function of scale factor and its derivative with respect to time. The critical term is the $Nm^2(c_1a^3+6c_2a^2+18c_3a+24c_4)$, which
   displays that $N$ is a Lagrangian  multiplier. In the Halmitonian form, this  multiplier ensures that the theory is ghost-free. One sees that this condition is exactly the same
   as what we obtained in the previous gauge in (\ref{constr}), which forbids a dynamical universe.

   A singular reference metric can evade the static universe problem. This is one of the key point in our study, and deserves to be demonstrated in a little more detailed way.  First of all, the ghost problem of the general theory of dRGT with singular reference metrics has been thoroughly discussed in \cite{self1, cao1}. In these previous works, the
 dRGT massive gravity with singular reference metrics is demonstrated to be ghost-free.

 Now we make a concise review of the discussion in \cite{self1}, and show how to apply it in the scenario of cosmology.
  The action of the dRGT massive gravity reads,
  \be
  S=\frac{1}{2\kappa^2}\int d^4x \left(\pi^{ij}\dot{\gamma}_{ij}+N_{\mu}R^{\mu}+V(N_{\mu},\gamma_{ij},f)\right),
   \label{actionmass}
    \en
    where  $N$ denotes the lapse and $N_i$ denote the shift functions in an ADM decomposition, $\gamma_{ij}$ is the spatial metric, $\pi^{ij}$ is the conjugate variables of $\gamma_{ij}$, and $V$
    is the dRGT potential, which has been shown in equation (4).  We define $N_{\mu}=(N,~N_i)$, and
 \be
 R^0=\sqrt{\gamma} \left[\textbf{R}+\frac{1}{\gamma}(\frac{\pi^2}{2}-\pi_{ij}\pi^{ij})\right],
 \label{hami}
 \en
 \be
 R^i=2\sqrt{\gamma}~\nabla_j\left(\frac{\pi^{ij}}{\sqrt{\gamma}}\right),
 \label{shami}
 \en
  where $\gamma$ denotes the determinant of $\gamma_{ij}$,  $\textbf{R}$ is the three dimensional Ricci scalar yielded by $\gamma_{ij}$.
    Apparently, all $N_{\mu}$ are no longer lagrange multipliers, and thus the Hamiltonian and momentum constraints are turned off.
       So all the six possible degrees of freedom of $\gamma_{ij}$ are liberated, including the ghost. The equations of motion of $R^{\mu}$ read,
       \be
       R^{\mu}+\frac{\partial V}{\partial N_{\mu}}=0.
       \label{pn}
       \en
      Thus, it is clear that $R^{\mu}$ are no longer constraints. This is the argument from Boulware-Deser, who claim that general non-linear massive gravity will be plagued by ghosts. If the Hamiltonian constraint
      is recovered, the ghost excitation will be killed. In  general case, one needs a transformation in  the parameter space in $(N, N_i)$ to find the desired Hamiltonian constraint. For detailed discussion of the
      transformation, see \cite{self1}.

      Here, in the scenario of cosmology, the problem becomes very simple. Considering the property of time-orthogonal of the FRW spacetime, one introduces
       in the ADM decomposition in the FRW metric,
       \be
       ds^2=-N^2(t)dt^2+a^2d\vec{x}^2,
       \label{FRWp}
       \en
        and for example the most simple singular reference metric $f_{\mu\nu}=$diag$(0,1,1,1)$ as the first case we discussed in above text. Then one directly obtain the potential,
        \be
 V=m^2\left(\frac{3c_1}{a}+ \frac{6c_2}{a^2} +\frac{6c_3}{a^3}\right).
      \label{poten3}
      \en
           The critical fact is that $N$ does not appear in this potential. Thus, from the equation of motion of $R_{\mu}$, as shown in (\ref{pn}), one derives,
           \be
           R^0=0.
           \label{hami}
           \en
           This is exactly the Hamiltonian constraint, which suppresses the ghost excitation.  The discussions of the ghost problem of the other
           cases of singular reference metrics just mimic this one. One sees that the problem is greatly simplified in the scenario of cosmology. Here the Hamiltonian constraint emerges just because of the lack of $f_{00}$. For more details of the general case of the theory with singular reference metrics, see \cite{self1, cao1}.

           As a comparison,  in the dRGT cosmology with a full-rank Minkowski reference metric, the potential becomes,
             \be
 V=m^2\left(\frac{3c_1}{a}+ \frac{6c_2}{a^2} +\frac{6c_3}{a^3}\right)+NF(a),
      \label{poten4}
      \en
      where $F(a)$ is given in (\ref{multiplier}),
      \be
      F(a)=m^2(c_1a^3+6c_2a^2+18c_3a+24c_4).
      \en
         So the Hamiltonian constraint (\ref{hami}) implies $F(a)=0$ by using (\ref{pn}). From this comparison one sees that in the case of singular reference metric, the Hamiltonian constraint recovers without any surplus constraint on $a$.

  One can make very similar analyses for a universe with spherical 3 subspace.  The situation of a universe with hyperbolic 3 subspace seems a little different, see \cite{lin}.

   Because of the inherent difficulties to realize a dynamical universe in dRGT the research interest in massive gravity gets decreased. One way to to get out of this tight corner is to consider a more complex theory. More fields, which are non-minimally coupled with gravity, had to be added
   to recover FRW cosmology \cite{morefields}. These approaches all involve more degrees of freedom. Inspired by the galilean theory, a galilean-like massive gravity is proposed to realize a dynamical universe \cite{galileon-like}. It is a hybrid theory of galilean and massive gravity. Giving up the Lorentz symmetry, a so-called ``minimal massive gravity" is suggested, which permits a homogeneous and isotropic
   universe \cite{minimal}. Even with the extra fields, the homogeneous and isotropic universe may be sill in absence in these extended massive gravities \cite{stillfail}.

   Let's scrutinize (\ref{poten2}) to find the reason why the universe cannot evolve. The equation of motion of $u$ imposes the constraint on $a$. One may think that if $u$ is not
  a field but a constant then it does not need an equation of motion. This assumption leads to the case of unitary gauge of the  Stuckelberg fields. We have seen that the same constraint
  appears in the unitary gauge. If $\dot{u}$ does not appear in (\ref{poten2}), the corresponding constraint (\ref{constr}) will vanish.  Based on this observation, an essential and straightforward
  method is to set $f_{00}=0$, which is a much more economic approach, without introducing any more freedoms. It is easy to confirm that the universe can be dynamical under this condition $f_{00}=0$.
  Similarly,  if one works in the unitary gauge but set $\phi^0=$constant, the last term in (\ref{multiplier}) vanishes. $N$ is no longer a Lagrangian multiplier, and thus the constraint (\ref{constr})
  vanishes spontaneously. $f_{00}=0$ may lead to a singular reference metric.

  A singular reference metric is not so weird and unacceptable as the first sight. In principle a reference metric has no direct relation with observables. Generally, a singular reference metric does not lead to physical difficulties.  In fact, besides cosmological considerations in the above text, there are several significant physical motivations to invoke a singular reference metric in massive gravity theory, for example, the AdS/CFT correspondence \cite{vegh}\cite{hu1} and neutron star and white dwarf structure \cite{nedw} etc.  A gauge-fixed massive gravity, for example a fixed reference metric diag$(-1,1,1,1)$, loses the property of diffeomorphism invariance.  In the scenario of AdS/CFT, the stress energy of corresponding field theory on the boundary is no longer conserved. Moreover, some special singular reference metric yields weak broken of the stress-energy, i.e., some components of the stress-energy conserved while the rest components dissipated. In fact, we need such a gravitational system to study the normal conductors in the scenario of AdS/CFT, where the momentums of the electrons dissipate when interacting with the host lattice. In this case, a fixed reference metric $\sim$ diag$(0,0,1,1)$ is required in studying the normal conductors \cite{vegh}. The stability problem, which cannot derived from the stability of dRGT, has been investigated in \cite{self1, cao1}. Massive gravity with singular reference metrics has been studied in several different aspects \cite{appl}. In cosmology one has seen that a singular reference metric may  significantly improve
  the massive cosmology by evading the extra constraint (\ref{constr}) on the scale factor. In the next section, we will study the massive cosmology with singular reference metric in details. One will see that a sound cosmology emerges. And furthermore, dark energy and dark matter appear naturally without any exotic matters.

  \section{massive cosmology with singular reference metrics}
  The general equation of motion of the physical metric $g$ corresponding to the action (\ref{action}) reads,
   \begin{eqnarray}
R_{\mu\nu}-\frac{1}{2}Rg_{\mu\nu}+m^2 \chi_{\mu\nu}&=&8\pi G T_{\mu \nu },~~
\label{field}
\end{eqnarray}
where $T_{\mu\nu}$ is the matter term corresponding to ${\cal L}_m$, $\chi _{\mu\nu}$ is the potential term of the gravity field,
\begin{eqnarray}
&& \chi_{\mu\nu}=-\frac{c_1}{2}({\cal U}_1g_{\mu\nu}-{\cal K}_{\mu\nu})-\frac{c_2}{2}({\cal U}_2g_{\mu\nu}-2{\cal U}_1{\cal K}_{\mu\nu}+2{\cal K}^2_{\mu\nu})
-\frac{c_3}{2}({\cal U}_3g_{\mu\nu}-3{\cal U}_2{\cal K}_{\mu\nu}\nonumber \\
&&~~~~~~~~~ +6{\cal U}_1{\cal K}^2_{\mu\nu}-6{\cal K}^3_{\mu\nu})
-\frac{c_4}{2}({\cal U}_4g_{\mu\nu}-4{\cal U}_3{\cal K}_{\mu\nu}+12{\cal U}_2{\cal K}^2_{\mu\nu}-24{\cal U}_1{\cal K}^3_{\mu\nu}+24{\cal K}^4_{\mu\nu}).
\end{eqnarray}

  \subsection{the static reference metric}
   Based on the previous discussions, we first set the reference metric $f_{\mu\nu}=$diag$(0,1,1,1)$. There is no dynamical variables in the reference metric. Thus in principle we should introduce the lapse function in the
  physical metric $g$.  As we analysed before, since $f_{00}=0$, the lapse function is no longer a Lagrangian multiplier in the action. An obvious result is that a dynamical universe becomes possible. Then it is unnecessary to introduce  the lapse function $N$ in the physical metric. We set the physical metric,
  \be
  ds^2=-dt^2+a^2d\vec{x}^2,
  \en
 where $d\vec{x}^2$ can be flat, spherical or hyperbolic, characteristiced by the spatial curvature $k=0,+1,-1$. Straightforward calculation presents $K_{\mu}^\nu=$diag$(0,1,1,1)/a$, and,
  \begin{eqnarray}
\label{calU}
&& {\cal U}_1= \frac{3}{a}, \nonumber \\
&& {\cal U}_2= \frac{6}{a^2}, \nonumber \\
&& {\cal U}_3= \frac{6}{a^3}, \nonumber \\
&& {\cal U}_4= 0.
\end{eqnarray}
  Then one derives the potential
 \be
 V=m^2\left(\frac{3c_1}{a}+ \frac{6c_2}{a^2} +\frac{6c_3}{a^3}\right),
      \label{poten3}
      \en
 which invokes no variables other than the scale factor $a$.

 We assume the matter, as usual, to be a perfect fluid,
 \be
  T_{\mu\nu}=\rho u_\mu u_\nu +p (u_\mu u_\nu+g_{\mu\nu}),
  \en
 in which $\rho$ and $p$ denote the density and pressure of the cosmic fluid, respectively. The field equation (\ref{field}) presents the Friedmann equations,
 \bea
 && H^2+\frac{k}{a^2}=\frac{8\pi G}{3}\rho -m^2\left(\frac{c_1}{2a}+\frac{c_2}{a^2}+\frac{c_3}{a^3}\right), \\
 && \frac{\ddot{a}}{a}=-\frac{4\pi G}{3}(\rho+3p)+m^2\left(-\frac{c_1}{4a}+\frac{c_3}{2a^3}\right).
 \ena
 From the Friedmann equations, one draws the effective density and pressure of the gravitons,
 \bea
 && \rho_g=-\frac{3m^2}{8\pi G}\left(\frac{c_1}{2a}+\frac{c_2}{a^2}+\frac{c_3}{a^3}\right),
 \label{rhog}
 \\
 && p_g=\frac{m^2}{8\pi G}\left(\frac{c_1}{a}+\frac{c_2}{a^2}\right).
 \label{pg}
 \ena



 It deserves to explicate the density and pressure of the gravitons in the above equations. It is well known that the stress energy of gravity field is an intricate problem \cite{self2}. In general relativity, a stress energy density is elusive, where only a kinetic term of the graviton, the Einstein-Hilbert term, is involved. Here the potential term of graviton is introduced.  In the case of massive gravity, the effective density and pressure in (\ref{rhog}) and (\ref{pg}) completely emerge from the potential term of the gravitons. The contributions of the kinetic term are not included. Therefore, (\ref{rhog}) and (\ref{pg}) do not imply that we have localized the stress energy of gravity fields. (\ref{rhog}) and (\ref{pg}) present the density and pressure of gravitons only in analogy to a perfect fluid, though they play the same role in the evolutions of the universe. This is sometimes called ``Einstein interpretation" of a modified gravity.

  Before discussing special cases of the Friedmann equations, we make a generic exploration of the massive cosmology. From (\ref{rhog}) and (\ref{pg}),
 \be
 \dot{\rho_g}+3H(\rho_g+p_g)=0.
 \en
 For a pure thermodynamic method to derive this equation, see \cite{self3}. So the gravitons and matters evolve independently in the history of the universe. First, we explore the ground state of such a universe, i.e., an empty universe without matters. In this case $\rho=p=0$.
 The equation of state of the effective stress energy of graviton reads,
 \be
 w_g=\frac{p_g}{\rho_g}=-\frac{1}{3}\frac{c_1/a+c_2/a^2}{c_1/(2a)+c_2/a^2+c_3/a^3}.
 \en
 This equation of state contains rich structure for the massive cosmology. Here we just present an example of an evolution of $w_g$ in the history of the universe.

 \begin{figure}
 \centering
 {\includegraphics[width=3.7in]{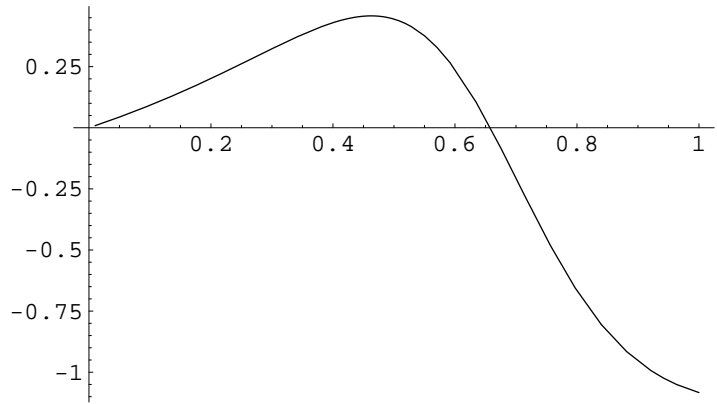}}
 \caption{ Effective EOS of the gravitons vs scale factor. This figure shows the evolution of the equation of state of the gravitons in an empty universe in massive gravity.}
 \label{EOS1}
 \end{figure}

 \begin{figure}
 \centering
 {\includegraphics[width=3.7in]{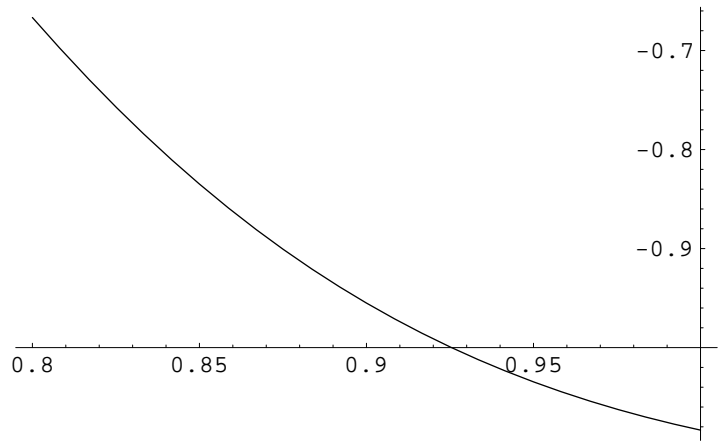}}
 \caption{Low redshift sector of fig 1. This figure clearly shows the behavior of crossing $w=-1$ of the equation of state of the gravitons, which is not a trivial issue \cite{self4}.     }
 \label{EOS2}
 \end{figure}

 In  fig 1 and fig 2, $c_1=-3.8,~c_2=2.5,~c_3=-1$. One sees that in the early universe the gravitons behave like a stiff matter, which is helpful for the structure formation, while becomes more and more softer in the later universe, which may be meaningful for the late time acceleration. $w_g$ crosses the phantom divide
 in some recent stage. It is clear that a dynamical universe is possible in massive gravity, though only a situation of empty universe is considered.
 \subsection{the dynamical reference metric}
  Further, we can set a dynamical reference metric, as done in the previous works \cite{drgtcos, lin}.  As a natural extension of the last subsection, we set $f_{\mu\nu}=b^2(t)$diag$(0,1,1,1)$. We call $b$ the scale factor of the reference metric (SFR). The invariant potential in the action (\ref{action}) reads,
  \be
      \sqrt{-g}~V= 3m^2\left(c_1ba^2+2c_2b^2a+2c_3b^3\right).
            \en
            $b$ does not appear in the other terms in the action. To reduce the arbitrariness of $b$, we consider an on-shell $b$. Thus a variation with respect to $b$ presents,
            \be
            c_1a^2+4c_2ba+6c_3b^2=0.
            \en
            So, if the reference metric is dynamical, the SFR $b$ is always proportional to the physical scale factor. The exact expression is,
            \be
            b=\frac{-2c_2\pm \sqrt{4c_2^2-6c_1c_3}}{6c_3}~a.
            \label{relationba}
            \en
          If one does not discuss complex metric, the present equation requires,
          \be
          2c_2^2\ge 3c_1c_3.
          \en
         This is an extra constraint required by cosmology, which does not appear in the general massive gravity theory. Define,
         \be
         B_{\pm}= \frac{-2c_2\pm \sqrt{4c_2^2-6c_1c_3}}{6c_3}.
         \en
         Then we write (\ref{relationba}) as $b=B_{\pm}a$. The Friedmann equations become,
         \bea
         \label{fried1}
          && H^2+\frac{k}{a^2}=\frac{8\pi G}{3}\rho -m^2\left(\frac{c_1B_{\pm}}{2}+{c_2B_{\pm}^2}+{c_3B_{\pm}^3}\right), \\
 && \frac{\ddot{a}}{a}=-\frac{4\pi G}{3}(\rho+3p)+m^2\left(-\frac{c_1B_{\pm}}{4}+\frac{c_3B_{\pm}^3}{2}\right).
         \ena
      The corresponding density of pressure of gravitons become,
          \bea
 && \rho_g=-\frac{3m^2}{8\pi G}\left(\frac{c_1B_{\pm}}{2}+{c_2B_{\pm}^2}+{c_3B_{\pm}^3}\right),
 \label{rhog}
 \\
 && p_g=\frac{m^2}{8\pi G}\left({c_1B_{\pm}}+{c_2B_{\pm}^2}\right).
 \label{pg}
 \ena
   Both $\rho_g$ and $p_g$ are constant, but,
   \be
   \dot{\rho_g}+3H(\rho_g+p_g)\neq 0,
   \en
      which implies that the gravitons are not adiabatic. With such a reference metric, an empty universe does not exist.
      It is similar to an interacting dark sectors model. One sees that, unexpectedly,  the case of dynamical reference metric $f_{\mu\nu}=b^2(t)$diag$(0,1,1,1)$ is not a simple generalization of the case of constant reference metric $f_{\mu\nu}=$diag$(0,1,1,1)$. The essence of the problem is that one introduce a new freedom $b(t)$, meanwhile its behavior gets restrict constraint from the on-shell equation $b\sim a$. Thus it cannot degenerate to the case of $b=$constant, only if the universe is static.

      We make a preliminary study of the dynamics of such a universe with dynamical reference metric. For simplicity, we consider a dust universe. The continuity equations for dust and gravitons can be written as,
      \bea
      \label{conti1}
      && \dot{\rho_d}+3H\rho_d=\Gamma,\\
      &&  \dot{\rho_g}+3H(\rho_g+p_g)=-\Gamma,
      \ena
     where $\Gamma$ denotes the energy flow between different sectors, which reads,
     \be
     \Gamma=\frac{3m^2H}{8\pi G}\left(\frac{c_1B_{\pm}}{2}+2c_2B_{\pm}^2+3c_3B_{\pm}^3\right).
     \en
    A positive $\Gamma$ denotes an energy flow from gravitons to the dust sector.
    Then we rewrite (\ref{conti1}) into,
    \be
    \label{conti2}
    a\frac{d\rho}{da}+3\rho=\frac{3m^2}{8\pi G}\left(\frac{c_1B_{\pm}}{2}+2c_2B_{\pm}^2+3c_3B_{\pm}^3\right).
    \en
    The solution is,
    \be
    \rho=C_1+\frac{{C_2}}{a^3},
    \en
    where
    \be
    C_1=\frac{m^2}{8\pi G}\left(\frac{c_1B_{\pm}}{2}+2c_2B_{\pm}^2+3c_3B_{\pm}^3\right),
    \en
    and $C_2$ is an integration constant.
   Rewriting the Friedman equation (\ref{fried1}) into,
   \be
   \frac{H^2}{H_0^2}=\frac{\Omega_{k0}}{a^2}+\frac{\Omega_{m1}}{a^3}+{\Omega_{\Gamma}}+{\Omega_{g}},
   \label{fried3}
   \en
  where,
  \be
  \Omega_{k0}=\frac{-k}{H_0^2},~\Omega_{m1}=\frac{8\pi G{C_2}}{3H_0^2},~\Omega_{\Gamma}=\frac{8\pi GC_1}{3H_0^2},~\Omega_{g}=\frac{8\pi G\rho_g}{3H_0^2}.
  \en
  When $\Omega_{k0}=0$, we find an analytical solution of (\ref{fried3}),
  \be
  a=2^{-4/3}\left({\Omega_{\Gamma}}+{\Omega_{g}}\right)^{-1/3}e^{-\sqrt{{\Omega_{\Gamma}}+{\Omega_{g}}}~H_0t+D}
  \left(e^{3\sqrt{{\Omega_{\Gamma}}+{\Omega_{g}}}~H_0t-D}-4\Omega_{m1}\right)^{2/3},
  \en
  where $D$ is an integration constant. When $e^{-D}=4\Omega_{m1}$, on can confirm that the scale factor reduces to the Einstein case,
  \be
  a\sim t^{2/3},
  \en
  at the limit ${\Omega_{\Gamma}}+{\Omega_{g}}\to 0$.

 Next we consider a radiation universe. In this case the Friedmann equation (\ref{fried3}) becomes,
 \be
   \frac{H^2}{H_0^2}=\frac{\Omega_{k0}}{a^2}+\frac{\Omega_{r0}}{a^4}+{\Omega_{\Gamma}}+{\Omega_{g}},
   \label{fried3}
   \en
   where $\Omega_{r0}$ is the relative composition of radiation at $a=1$. We find the exact solution of the
   above equation,
   \be
   a=\frac{1}{2}\sqrt{\frac{e^{2\sqrt{c_1}H_0t+D}-4\Omega_{r0}e^{-2\sqrt{c_1}H_0t-D}}{\sqrt{c_1}}}~,
   \en
   where $D$ is the integration constant. When $e^D=2\Omega_{r0}^{1/2}$, and $c_1\to 0$, $a$ comes back to the
   radiation universe $a\sim t^{1/2}$ in GR.


   \subsection{more complicate reference metric}
   One has seen that massive cosmology leads to ordinary dynamical cosmology with some special reference metrics. The two resulted models are
   familiar in the studies of cosmology. Now we develop a different one in massive cosmology. The reference metric is subtle in massive gravity.
   It is necessary for endowing mass for a graviton. However, it blocks some ``obvious results" in general relativity, for example, an FRW universe,
   a \sch~like black hole, and Kerr-like rotating one. In all these cases, one has to invoke  non-trivial reference metrics \cite{liping}.
    Generally, the physical motivations of these reference metrics are not very clear. We take such reference metric largely due to the wanted results.   Now, let's
   introduce a hybrid reference metric in the previous two subsections, $f_{\mu\nu}=$diag$(0,1,b^2,b^2)$. Then one reaches, ${\cal K}_{\mu}^{\nu}=$diag
   $(0,1/a, b/a, b/a)$. Under this reference metric, ${\cal U}_i$ read,
   \begin{eqnarray}
\label{calU1}
&& {\cal U}_1= \frac{1+2b}{a}, \nonumber \\
&& {\cal U}_2= \frac{4b+2b^2}{a^2}, \nonumber \\
&& {\cal U}_3= \frac{6b^2}{a^3}, \nonumber \\
&& {\cal U}_4= 0.
\end{eqnarray}
  Terms in the action related to $b$ read,
  \be
  \sqrt{-g}~V= 3m^2\left(c_1a^2(1+2b)+2c_2a(2b+b^2)+6c_3b^2\right).
            \en
  Similar to the previous case, we consider an on-shell $b$. A variation with respect to $b$ requires,
  \be
  b=-\frac{c_1a^2+2c_2a}{2c_2a+6c_3}.
  \label{baa}
  \en
 The Friedmann equations in this case,
 \bea
 \label{fried2}
 && H^2+\frac{k}{a^2}=\frac{8\pi G}{3}\rho -m^2\left(\frac{c_1(1+2b)}{6a}+\frac{c_2(2b+b^2)}{3a^2}+\frac{c_3b^2}{a^3}\right), \\
 && \frac{\ddot{a}}{a}=-\frac{4\pi G}{3}(\rho+3p)+m^2\left(\frac{c_1(1-4b)}{12a}+\frac{c_2(b-b^2)}{3a^2}+\frac{c_3b^2}{2a^3}\right).
 \ena
 One extracts the effective density and pressure of gravitons from the above equations,
 \bea
 && \rho_g=-\frac{3m^2}{8\pi G}\left(\frac{c_1(1+2b)}{6a}+\frac{c_2(2b+b^2)}{3a^2}+\frac{c_3b^2}{a^3}\right),
 \label{rhog2}
 \\
 && p_g=\frac{m^2}{8\pi G}\left(\frac{c_1b}{a}+\frac{c_2b^2}{a^2}\right),
 \label{pg3}
 \ena
  where $b$ is given by (\ref{baa}). Before considering the full complexity of the above equations, we study some special cases. First, when $c_1=c_3=0$, the gravitons becomes
  stiff matters with $w=1$ Then, when $c_2=0$, one obtains $b\sim a^2$. Thus from (\ref{rhog2}) the density of the gravitons increases when the universe expands.

 In the case of $c_1\neq 0$ $c_2\neq 0$ $c_3\neq 0$, the evolution of the universe with gravitons with reference metric $f_{\mu\nu}=$diag$(0,1,b^2,b^2)$ is fairly complex.
 To understand the evolution of the universe with massive gravitons, we first examine the energy conservation of the gravitons,
 \be
   \dot{\rho_g}+3H(\rho_g+p_g)=\frac{m^2}{8\pi G} \frac{(2c_2^2-3c_1c_3)(6c_3+4c_2a+c_1a^2)\dot{a}}{2a^2(3c_3+c_2a)^2}.
   \label{EF}
   \en
  Generally speaking the stress-energy of the massive gravitons is not conserved. Thus it needs the other components in the universe to undertake energy flows. A special case is $\alpha=2c_2^2-3c_1c_3=0$, in which the gravitons evolve adiabatically. We call the massive gravity with such parameters critical massive gravity. In the critical massive gravity, the ground state of the universe, i.e., an empty universe, exists,
  \be
  a=C_1\exp\left[\left(\frac{c_1^3}{24c_3}\right)^{\frac{1}{4}}t\right],~~a=C_2\exp\left[-\left(\frac{c_1^3}{24c_3}\right)^{\frac{1}{4}}t\right],
  \label{desi}
  \en
    which is an analogy to the Minkowski space in general relativity.  It is easy to recognize that the ground state described by (\ref{desi}) is de Sitter space. It can be an expanding or contracting universe. A note is that
    a superposition of an expanding and contracting universe is not a solution of the field equation because of its non-linearity.

    To more realistically describe the evolution of the universe at some late time, we introduce a dust component. In this case, (\ref{fried2}) becomes,
    \be
    \label{fried3}
    \frac{H^2}{H_0^2}=\frac{\Omega_{k0}}{a^2}+\frac{\Omega_{m0}}{a^3} -\frac{m^2}{H_0^2}\left(\frac{c_1(1+2b)}{6a}+\frac{c_2(2b+b^2)}{3a^2}+\frac{c_3b^2}{a^3}\right).
    \en
    The solution of the above equation reads,
    \be
    a=L\frac{\tanh^{2/3} {(Mt+C_3)}}{\left(1-\tanh^2 {(Mt+C_3)}\right)^{1/3}},
    \en
    where,
    \be
    L=\frac{\sqrt{2}(3\bar{c}_3)^{1/6}\Omega_{m0}^{1/3}}{\sqrt{\bar{c}_1}},~~~M=\frac{2^{1/4}3^{3/4}\bar{c}_1^{3/4}}{4\bar{c}_3^{1/4}},
    \en
   and $C_3$ is an integration constant. $\bar{c}_1$ and $\bar{c}_3$ are defined as,
   \be
   \bar{c}_1=\frac{m^2}{H_0^2}c_1,
   \en
   and
   \be
   \bar{c}_3=\frac{m^2}{H_0^2}c_3,
   \en
   respectively.

  Now we deal with the non-critical case $\alpha\neq 0$.
    We assume that the dark matter balances the energy flow from the gravitons (\ref{EF}),
    \be
    \dot{\rho_m}+3H\rho_m=-\frac{m^2}{8\pi G} \frac{(2c_2^2-3c_1c_3)(6c_3+4c_2a+c_1a^2)\dot{a}}{2a^2(3c_3+c_2a)^2}.
    \en
    When $c_2\neq0$, the  explicit form of $\rho_m$ reads,
  \be
  \rho_m=\frac{C_3}{a^3}-\frac{\xi\alpha}{4c_2^4a^3}\left(a^2c_1c_2^2-\frac{18\alpha c_3^2} {c_2a+3c_3}+4\alpha a-18\alpha c_3\log(c_2a+3c_3)\right),
  \label{density}
  \en
  where,
  \be
  \xi=\frac{m^2}{8\pi G},
  \en
  $C_3$ is an integration constant, and $\alpha$ is the critical parameter we defined before.
  Clearly, it degenerates to the non-interacting case when $\alpha=0$.
  $c_2=0$ is a singularity of the above solution. One has to deal with this case specially. When $c_2=0$, one obtains,
  \be
  \rho_m=\frac{C_4}{a^3}+\xi\left(\frac{c_1}{2a}+\frac{c_1^2a}{24c_3}\right).
  \en

  Now, as an example, we fit the case with $c_2\neq0$, which has the most rich evolution behaviors. We rewrite the Friedmann equation (\ref{fried3}) as,
  \be
    \label{fried5}
    \frac{H^2}{H_0^2}=\frac{\Omega_{1}}{a}
    +\frac{\Omega_{2}}{a^2}+\frac{\Omega_{3}}{a^3} +\Omega_{\lambda}+\frac{\Omega_{1p}}{a+3c_3/c_2}+\frac{3\alpha^2 c_3m^2}{2H_0^2c_2^4a^3}\ln (a+\frac{3c_3}{c_2}).
    \en
    Here,
    \be
    \Omega_{1}=\frac{m^2}{3H_0^2}\left(\frac{\alpha^2}{4c_2^2c_3}\right),
    \en
    \be
    \Omega_{2}=\Omega_{k0}-\frac{m^2}{3H_0^2}\left(\frac{\alpha^2}{2c_2^3}+\frac{\alpha^2}{c_2^4}\right),
    \label{ome2}
    \en
    \be
    \Omega_{3}=\Omega_{b0}+\frac{8\pi G}{3H_0^2}C_3+\frac{m^2}{3H_0^2}\frac{9\alpha^2c_3\ln c_2}{2c_2^4},
    \en
    \be
    \Omega_{\lambda}=\frac{m^2}{3H_0^2}\frac{c_1^2}{4c_2},
    \en
       \be
    \Omega_{1p}=\frac{m^2}{3H_0^2}\left(\frac{\alpha^2}{12c_2^2c_3}\right),
    \en
   and $\Omega_{b0}$ denotes the present component of baryonic matters, which does not interacts with gravitons.

    In  general, we set $\Omega_2=0$, and,
\begin{eqnarray}
&&\Omega_{1p0}=\frac{\Omega{1p}}{1+3c_3/c_2},\\
&&\Omega_{4}=\frac{3\alpha^2c_3}{2H_0^2c_2^4},\\
&&\Omega_{40}=\Omega_{4}\ln(1+3c_3/c_2).
\end{eqnarray}
    We stress that $\Omega_2=0$ does not implies a spatially flat universe, since the massive gravitons also contribute $\Omega_2$ significantly, from (\ref{ome2}).
    For any $k$ in a realistic universe, it behaves like a spatially flat one for proper parameters in massive gravity.

    Here,  we apply the Pantheon data \cite{Scolnic:2017caz}, CMBR data \cite{Ade:2015xua} and the BAO data \cite{Percival:2009xn,Blake:2011en,Beutler:2011hx,Giostri:2012ek} to contrain the model.
     The code of Cosmomc \cite{Lewis:2002ah} is used in this fitting. We set the range of  the parameters are  $ \Omega_m0=[0.005,1.0]$, $\Omega_10=[-0.3,0.3]$, $\Omega_{1p0}=[-0.2,0.4]$, $\Omega_{40}=[-0.1,0.2]$,  and $3c_3/c_2=[0.5,1]$.

\begin{figure}
  \centering
   {\includegraphics[width=3.7in]{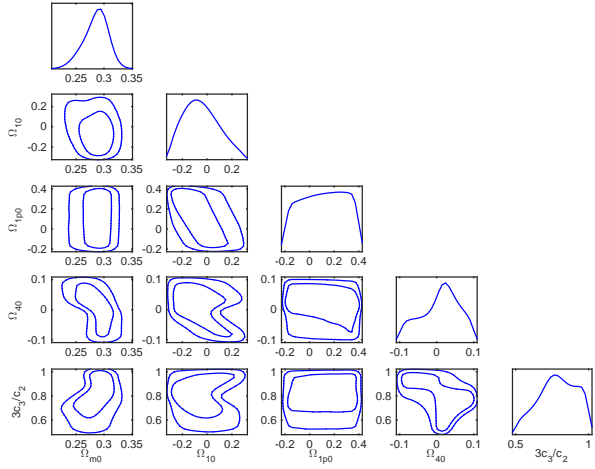}}
   \caption{One-D marginalizd distributions and two-D joint distributions for  the parameters with the $1\sigma$ and $2\sigma$ contours.}
  \label{masstri}
\end{figure}

The  constraining results present $ \Omega_{m0}=0.287_{-0.072-0.093}^{+0.049+0.066}$,  the best fit of other value are $\Omega_{10}=0.077$, $\Omega_{1p0}=0.357$, $\Omega_{40}=-0.079$,  $3c_3/c_2=0.952$ with their $1\sigma$ and $2\sigma$ range very close to the prior range. From fig. \ref{masstri}, one sees that the present observation leave enough space for the parameters.

   It is easy to see that $\Omega_{1}$ and $\Omega_{1p}$ vanish, and $\Omega_{2}$ and $\Omega_{3}$ reduce to the ordinary case under the condition $\alpha=0$. This is exactly the
   the critical case which we studied in the above context. The above equation cannot directly reduce to the case with $c_2=0$ or $c_3=0$. We discuss these cases separately. The Friedmann equation corresponding to $c_2=0$ reads,
  \be
    \label{fried6}
    \frac{H^2}{H_0^2}=\frac{\Omega_{k0}}{a^2}+\frac{\Omega_{3}}{a^3} +\Omega_{+1}{a}.
    \en
    Here,
    \be
    \Omega_{+1}=\frac{m^2}{3H_0^2}\frac{c_1^2}{8c_3},
    \en
    \be
    \Omega_{3}=\Omega_{b0}+\frac{8\pi G}{3H_0^2}C_4,
    \en
      and $c_3=0$ reads,
 \be
    \label{fried7}
    \frac{H^2}{H_0^2}=\frac{\Omega_{2}}{a^2}+\frac{\Omega_{3}}{a^3} +\Omega_{\lambda}.
    \en
    Here,
    \be
    \Omega_{2}=\Omega_{k0}+\frac{m^2}{3H_0^2}(-3c_2),
    \en
    \be
    \Omega_{3}=\Omega_{b0}+\frac{8\pi G}{3H_0^2}C_3,
    \en
    \be
    \Omega_{\lambda}=\frac{m^2}{3H_0^2}\frac{c_1^2}{4c_2}.
    \en
    These two cases are relatively simple. We do not further constrain them. From the above discussions, one sees that
    the massive gravity with singular reference metric permits dynamical universe with all three cases of spatial curvatures.

    \section{conclusion}
    The inexistence of a dynamical universe with negative and zero spatial curvatures is a critical difficulty in the studies of massive gravity.
    We demonstrate a straightforward and much more economic way to overcome this problem. After careful analysis of this problem, we find that
    a singular reference metric can remove the extra constraint in the equation of motion, and thus leads to a sound cosmology with all three types of spatial curvatures.

    We study three cases of singular reference metrics, which are static, dynamical, and a hybrid one. All of them permit  cosmology with all three types of spatial curvatures.
    We preliminarily constrain the last one with SNe, CMBR, and BAO data. The result shows that the observations leave enough space for the model parameters.

 {\bf Acknowledgments.}
   H.Z. thanks C. de Rham and A. Tolley for helpful discussions. This work is supported in part by the National Natural Science Foundation of China (NSFC) under grant Nos. 11575083, 11565017, Shandong Province Natural Science Foundation under grant No.ZR201709220395, CQ CSTC under grant No. cstc2015jcyjA00044 and CQUPT under grant No. A2009-16.

\end{document}